\documentclass[12pt]{JHEP}
\usepackage{epsfig}

\def\fun#1#2{\lower3.6pt\vbox{\baselineskip0pt\lineskip.9pt
  \ialign{$\mathsurround=0pt#1\hfil##\hfil$\crcr#2\crcr\sim\crcr}}}
\newcommand{\beq}{\begin{equation}}
\newcommand{\eeq}{\end{equation}}
\title{Magnetic lensing of Extremely High Energy\\
Cosmic Rays in a Galactic wind}

\author{Diego Harari$^a$, Silvia Mollerach$^b$ and Esteban Roulet$^b$
\\$^a$Departamento de F\'\i sica, FCEyN, Universidad de Buenos Aires
\\Ciudad Universitaria - Pab. 1, 1428, Buenos Aires, Argentina
\\$^b$Departamento de F\'\i sica, 
Universidad Nacional de La Plata\\ CC67,  1900, La Plata, Argentina
\\ Email: \email{harari@df.uba.ar, mollerac@venus.fisica.unlp.edu.ar, 
roulet@venus.fisica.unlp.edu.ar}}

\abstract{
We show that in the model of Galactic magnetic wind recently proposed
to explain the extremely high energy (EHE) cosmic rays 
so far observed as originating
from a single source (M87 in the Virgo cluster), the magnetic field 
strongly magnifies the fluxes and produces multiple images of the source. 
The apparent position on Earth of the principal image moves, 
for decreasing energies, towards the galactic south. 
It is typically amplified by an order of magnitude  
at $E/Z\sim 2\times 10^{20}$~eV, but becomes strongly demagnified 
below $10^{20}$~eV.
At energies below $E/Z\sim 1.3\times 10^{20}$~eV, all events in the
northern galactic hemisphere  are due to secondary images, which have
huge amplifications ($>10^2$).
This model would imply strong asymmetries between the north
and south galactic hemispheres, such as a (latitude dependent)
upper cut-off value below
$2\times 10^{20}$~eV for CR protons arriving to the south and 
lower fluxes in the south than in the north above
$10^{20}$~eV. The large resulting magnifications reduce the power
requirements on the source, but the model needs a significant tunning
between the direction to the source and the symmetry axis of the wind.
If more modest magnetic field strengths were assumed, 
a scenario in which the observed EHE events are heavier nuclei 
 whose flux is strongly lensed becomes also plausible 
and would predict that a transition from a light composition to a 
heavier one could take place at the highest energies.}

\preprint{.}

\begin{document}

\section{Introduction}

The origin and nature of the highest energy cosmic rays so far detected
stands as a puzzle for contemporary astrophysics \cite{bh98}. 
The flux of protons 
with energy around and above 70~EeV (1 EeV = $10^{18}$~eV) should be 
significantly attenuated over distances of order 100~Mpc due to their
interaction with the cosmic microwave background~\cite{GZK}. 
Nuclei should be attenuated over even shorter distances~\cite{pu76}.
Nonetheless, fourteen events with estimated energy above 100~EeV
have been detected so far, largely exceeding the
expected fluxes if the sources are at cosmological distances. 
This suggests a ``local'' origin of extremely high energy cosmic
rays (EHECRs). The puzzle arises because the angular distribution 
of the fourteen EHE events is consistent 
with isotropy (given the limited statistics and insufficient sky 
coverage) and there are no known sources near their arrival directions 
and inside our 100~Mpc neighborhood considered to be a potential site 
for acceleration of cosmic rays to such enormous energies.

A potential solution to the puzzle is that EHECRs are protons or nuclei
that do indeed originate in sources within a 100~Mpc neighborhood of 
Earth, but their arrival directions do not point to their place of 
origin because  their trajectories are significantly bent as they 
traverse intervening magnetic fields.
The regular component of the magnetic
field in the Milky Way leads to sizeable deflections~\cite{st97,me98}
and other magnetic lensing effects upon ultra high energy charged cosmic
rays, such as flux (de)magnification
and multiple image formation~\cite{I,II}. This is the case, however,
only if the ratio $E/Z$ between energy and electric charge of the CRs is
below approximately 30~EeV. The EHECRs would
thus be severely affected by the regular component of the galactic
magnetic field only if they have a significant component which is 
not light.

It has recently been speculated \cite{ahn99} that all the events so
far detected at energies above $10^{20}$ eV may originate from M87 in the
Virgo cluster, if the Galaxy has a rather strong and extended
magnetic wind. Indeed, such extreme galactic magnetic wind is 
compatible with an origin for all EHE events at less than $20^\circ$
from the direction to the Virgo cluster, if two out of the thirteen
events considered are He nuclei, the rest being protons.
One event was excluded from the dataset due to the large uncertainties 
in its energy determination. 

In this paper we further develop the analysis of the scenario
put forward in \cite{ahn99}. The determination of the deflections 
of CR trajectories is
insufficient to test the consistency of the scenario with the 
observational data and to determine its generic predictions,
because magnetic lensing also produces huge flux (de)magnification 
and multiple image formation, effects that we analyse here and which
turn out to be crucial to establish the detailed features of the model.

\section{Flux enhancement by magnetic lensing in a galactic wind}
\label{magnification}

We consider an azimuthal magnetic field with strength given by 
\begin{equation}
B=B_0{r_0\over r}\sin\theta \tanh(r/r_s)
\end{equation}
as a function of the radial (spherical) coordinate, $r$, and the angle
to the north galactic pole, $\theta$. The distance from Earth
to the galactic center is 
$r_0=8.5~$kpc and the local value of this wind field is taken as 
$B_0 = 7\mu$G. This field has the 
$\sin\theta/r$~dependence adopted in ref. \cite{ahn99} 
with an extra smoothing in
the Galactic center region, given by the $\tanh(r/r_s)$ factor,
introduced to avoid unphysical divergences at small radii. For
definiteness we took  $r_s=5$~kpc but the results are not dependent on
the precise way in which this smoothing is performed. We also
adopted a 1.5~Mpc cutoff for the extension of the field as in ref.
\cite{ahn99}.

The trajectories of CR protons in the Galactic wind magnetic field are
obtained by backtracking antiprotons leaving the Earth. 
The azimuthal nature of this field bends all the trajectories
towards the north galactic pole \cite{ahn99,bi00}.
 The incident direction outside the
region of influence of the wind points at less than $15^\circ$ from the
north pole for 
 all the observed EHE events, except for the two most energetic 
ones. The two most
energetic events could also come from that cone only if their
electric charge is assumed to be larger, 
for example if they are He nuclei. This fact has been
exploited to suggest that all the EHE events may have a common source,
M87 in the Virgo cluster \cite{ahn99}, which is indeed at
$b=74.4^\circ$, i.e. not far from the north galactic pole. Clearly at
this level one has to be satisfied  with this kind of accuracy in the
pointing since the wind model is highly idealised  and one is also
neglecting  the additional deflections that should take place near the
source and in the travel through the intergalactic medium. This will
also require that our conclusions be based only  on the general
qualitative  features of the model, rather than on its specific
details. 

\FIGURE{\epsfig{file=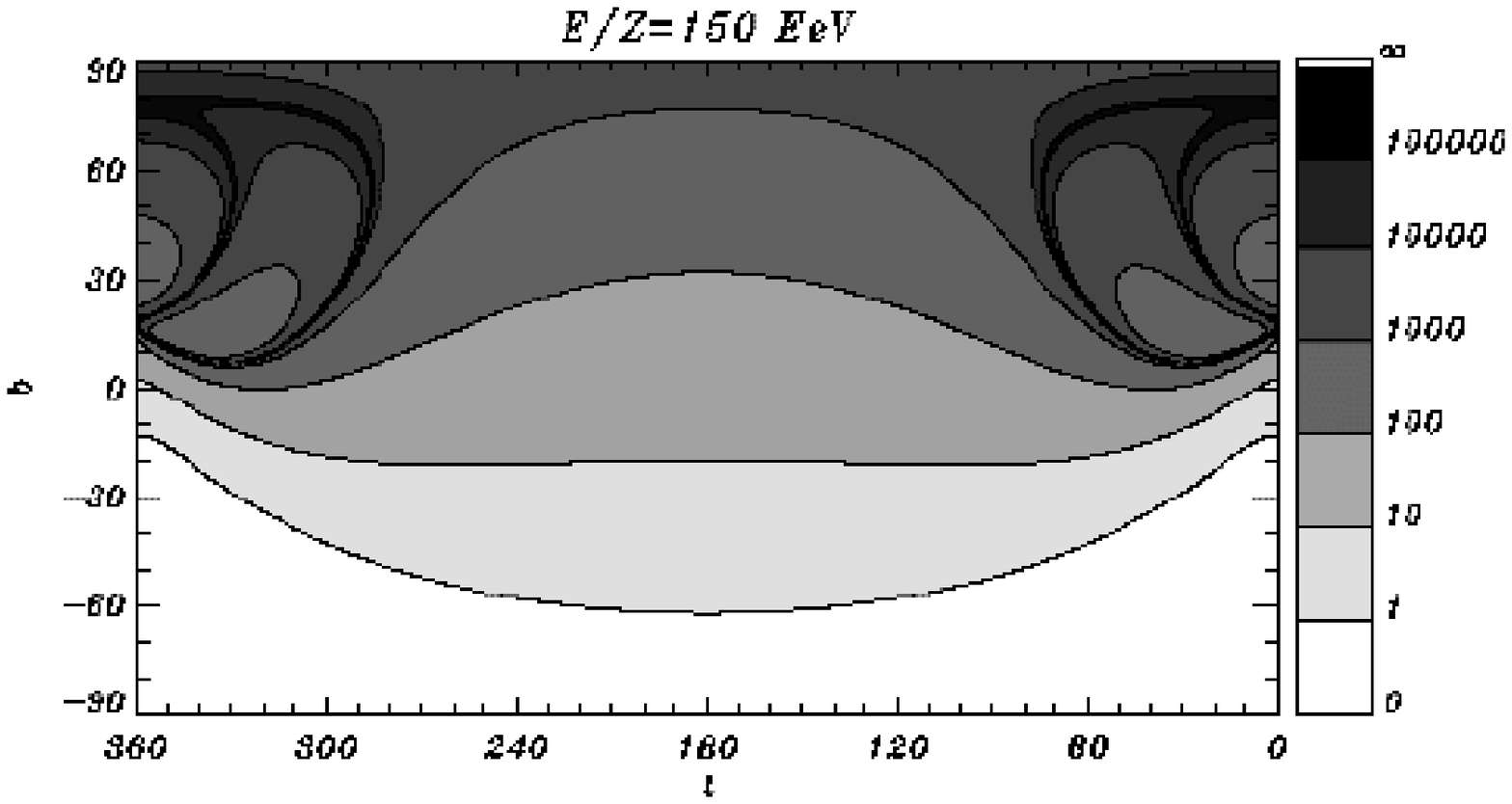,width=12cm}
\epsfig{file=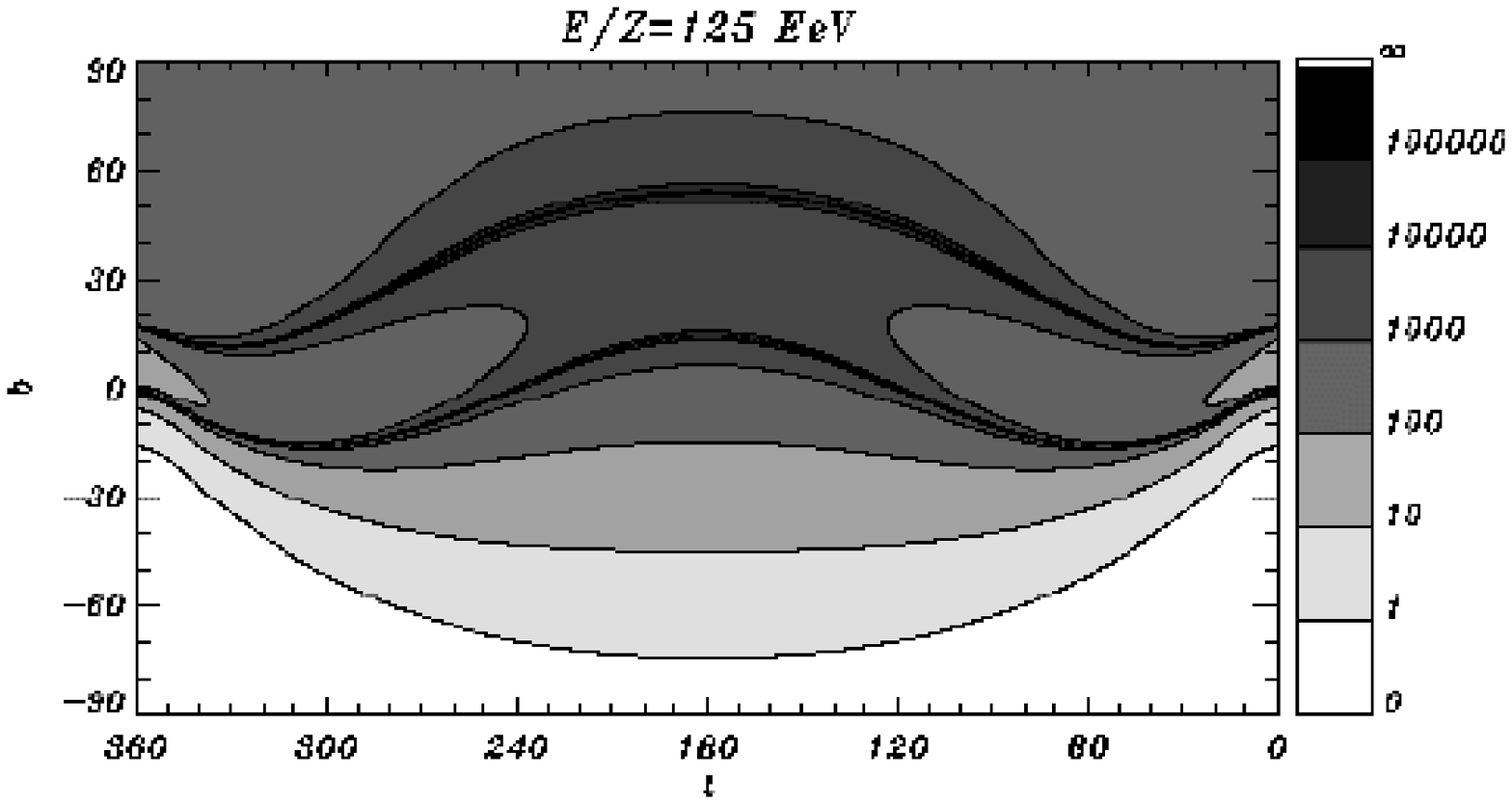,width=12cm}
\epsfig{file=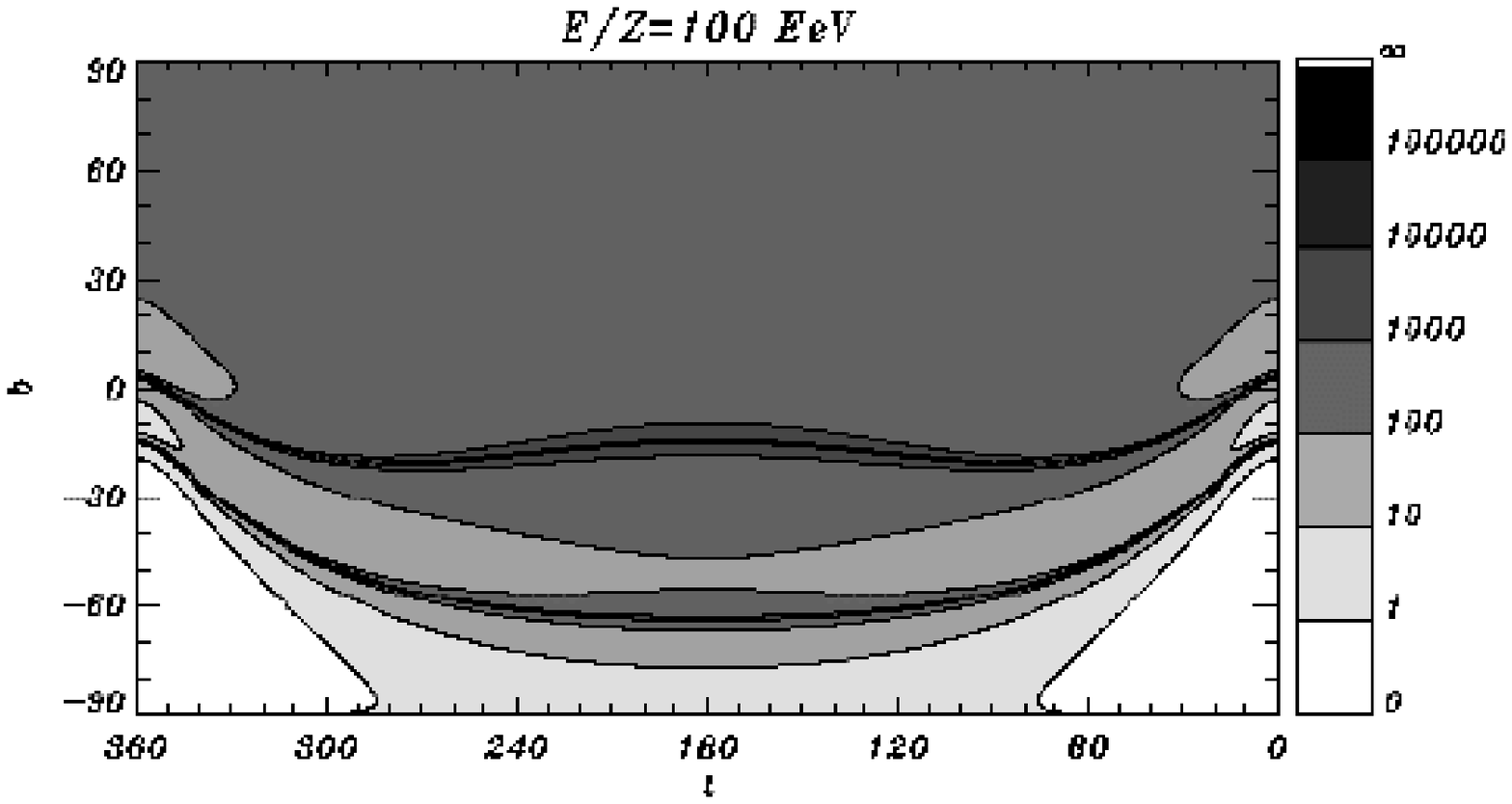,width=12cm}
\caption{Contour plots of
the magnification  due to 
the galactic magnetic wind model of the CR flux from a point
source. The plots are shown as a function of
the  arrival direction at the Earth and  for three different CR energies.}}

It has been pointed out that large scale magnetic fields  not only
deflect CR trajectories but that they  also act as a giant lens
(de)magnifying the fluxes received from different directions and
leading to multiple image formation \cite{I}.
For a given source this lensing effect varies with the energy and thus
can enhance or attenuate the fluxes differently for different energies. 
This effect turns out to be 
quite strong in the Galactic wind model under consideration, so 
it  has to be taken into account in a complete 
analysis of this scenario.

The magnification due to the lensing effect is
computed as the ratio between the area subtended by a parallel bundle
of particles arriving to the Galactic wind region and that subtended 
by the same particles at their arrival to the Earth \cite{I}. 
Fig. 1 shows the contour plots of equal magnification for three values
of the $E/Z$ ratio. Each point denotes the arrival direction of a 
CR to the Earth in galactic coordinates $(\ell,b)$. 
We observe that huge magnifications, in excess than a
factor of 100, are attained in large regions of the sky. This is the
case for most directions with $b > 15^\circ$ for $E/Z = 150$~EeV and
with $b > - 15^\circ$ for $E/Z = 125$~EeV. The critical lines (lines
where the magnification diverges and that correspond to the caustic
lines in the source plane) move quickly to the south as the energy
decreases. Most of the sky is swept by the critical lines as $E/Z$
varies between 150~EeV and just below 100~EeV.
Since the magnification is huge in the regions close to these
lines, a strong enhancement in the detection probability of events with
energies close to that at which the caustics cross the source
direction is expected.

\section{Multiple images}
\label{images}

\FIGURE{\epsfig{file=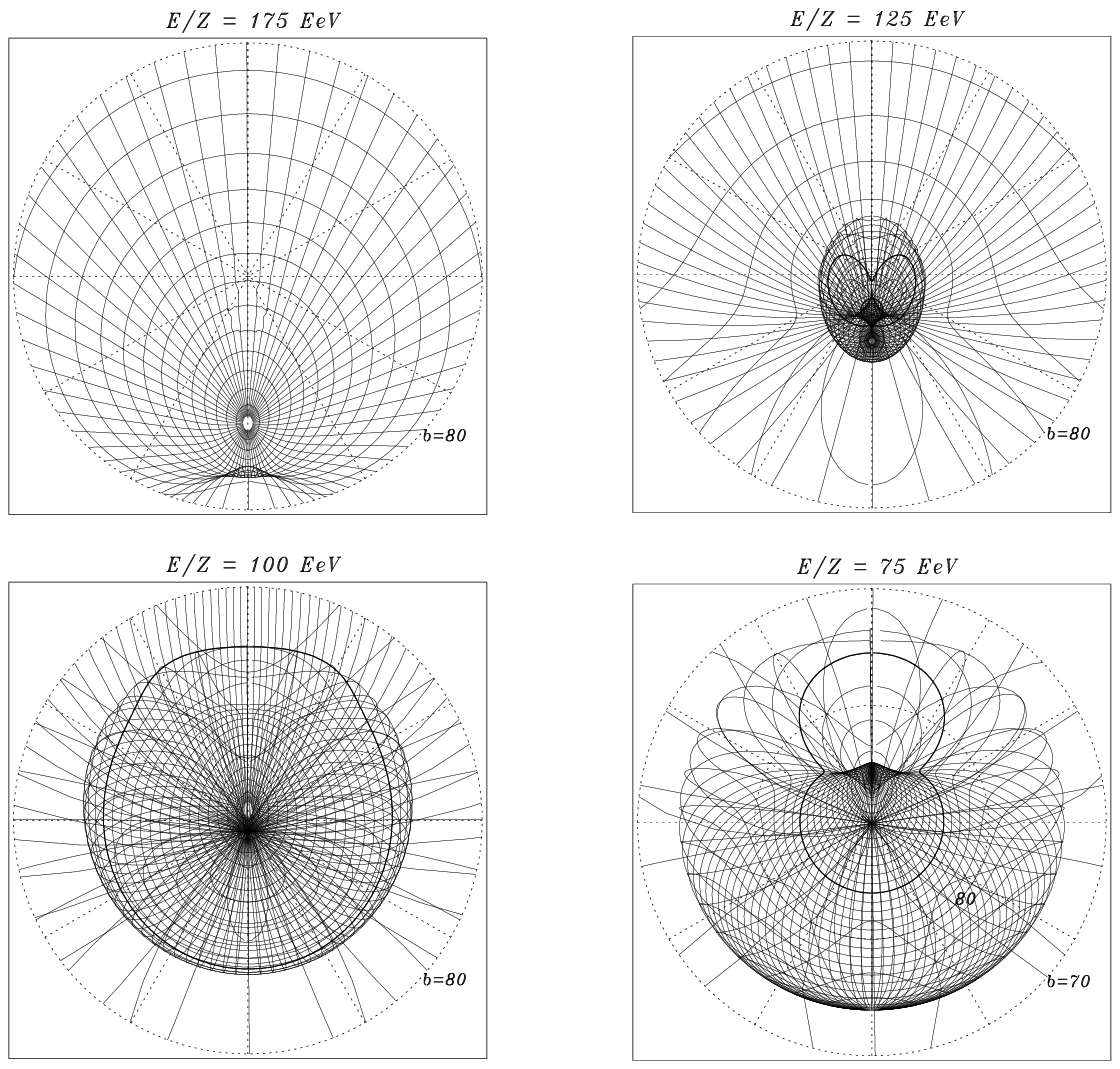,width=14.5cm}
\caption{Projection of  a regular grid
(in polar coordinates centered at the northern galactic pole)
of the sky seen on Earth onto the source sky, i.e. the corresponding 
 directions from which CRs enter the galactic 
wind.
Sources located in regions where this ``sky sheet''
 is folded have multiple
images. Three images are seen if the source is inside the ``blob''
(which subtends an angular diameter of the order of $5^\circ$ at
$E/Z=125$~EeV, $15^\circ$ at
$E/Z=100$~EeV, and $30^\circ$ for $E/Z=75$~EeV), 
five images if it is inside the much smaller 
``diamond''. Dashed lines represent a polar grid of source 
directions, spaced every $10^\circ$ in galactic 
latitude and 
$30^\circ$ in longitude. In the lowest energy case  the 
$b>70^\circ$ cap is plotted while in the rest just
the $b>80^\circ$ cap is shown in order to  better appreciate 
the details.}}

The presence of critical curves in the magnification maps indicates
that the magnetic lensing effect of the Galactic wind  produces
multiple images of the source for some source directions in the range
of energies considered. This fact is most clearly seen from the plots
of the ``sky sheets'', representing the projection of a regular grid of
observing directions at Earth to the directions outside of the Galactic
wind region. This kind of plots map the arrival direction(s) at which a
CR is observed at Earth to the direction from where the CR arrived to
the region of influence of the magnetic field.
Fig.~2 shows this sheet in polar coordinates centered in the
north pole for three values of the $E/Z$ ratio. The stretching of this
sky sheet reflects the magnification for those source
directions, very stretched regions being strongly demagnified and
those where the sky sheet is densely contracted being very
magnified. The fold lines correspond to the
position of the caustics, where the magnification diverges.
Sources located
in regions where the sheet is folded are observed at Earth from
several different directions. 
For $E/Z = 175$~EeV a couple of folds ending into two cusps (a lip)
develop. 
A source inside the folds has three different images.
Note however that the folds cover only a tiny fraction of the sky
and thus this is a rare effect at these energies. 
For $E/Z = 125$~EeV a complete blob has
developed on the sheet. The blob is connected to the
rest of the surface through a diamond shaped caustic (this is due to the
Earth eccentric position in the Galaxy: for observers in the symmetry
axis, e.g. at the Galactic
center, the diamond would shrink to a
point\footnote{These caustics are analogous to those associated to 
elliptical lenses, or spherical ones with external shear, 
in gravitational  lensing \cite{glsheets}, for which the central
diamond shrinks to a point-like caustic in the circularly symmetric
limit.}). The southernmost critical
line in the second panel of Fig. 1 corresponds to the diamond caustic
in Fig.~2, while the northern critical line corresponds to the
circular fold caustic. The huge amplifications found for all the
directions in the northern sky above  
 the southernmost critical line reflect the fact that a
large fraction of the sky seen on Earth 
 is shrinked inside the blob, and hence the sheet is highly contracted
there. While the
region inside the diamond caustic (leading to five images) is still
tiny in the source sky, the blob (leading to three images) 
covers a region of $\sim 5^\circ$ 
diameter. We have also drawn, with a bold solid line, the directions
for which the cosmic rays would arrive to the Earth along the galactic
equator. It is clear that all 
cosmic rays with $E/Z=125$~EeV with arrival directions
in the northern galactic hemisphere actually enter the galactic wind
from directions less than $2.5^\circ$ away from the center of the blob. 
At smaller energies a still larger fraction
of observing directions are swallowed into the blob (this corresponds
to the motion of the critical lines towards the south in Fig.~1),
which at $E/Z=100$~EeV has an angular diameter of the order of 
$15^\circ$. At $E/Z=75$~EeV the
diamond has nearly disappeared, and a new blob starts developing on 
top of the previous blob. At this energy the angular diameter of the
main blob, corresponding to the source positions leading to multiple
images, is already larger than $30^\circ$.

It is important to notice that only if the blob is on top of the
source position (eventually displaced by the energy dependent
deflections due to the extragalactic magnetic fields) the cosmic rays
are able to reach the Earth along directions above the southernmost
critical line in Fig.~1, which for $E/Z=125$~EeV already covers the
whole  northern hemisphere. Hence, protons  with $E<125$~EeV can only
reach the northern hemisphere as extremely magnified ($A>10^2$) 
secondary images.

\section{Generic predictions}
\label{predictions}

We now discuss some generic features of 
the scenario in which all the CR events 
with energies above $10^{20}$~eV so far detected 
originate from M87 in the Virgo cluster, their trajectories having 
been significantly bent by this rather strong and extended galactic
magnetic wind \cite{ahn99}. These generic properties may serve to 
test the validity of the scenario as the data on EHECRs increase.

Detailed predictions depend upon the exact nature and strength
of the galactic wind, as well as upon the precise deflections suffered by
the CR trajectories from the source in M87 up to their entrance to
the galactic wind. Nevertheless, the generic features of any 
scenario compatible with current observations, mainly determined by 
the focusing properties of the magnetic wind, should be similar to 
those in the highly idealised model analysed here. 

In the model under consideration, all thirteen
EHECR events enter the galactic wind region along different
directions, most of them just between $2^\circ$ and $5^\circ$ away 
from the direction to the north galactic pole, and at most 
separated by $12^\circ$ from the vertical direction. 
Intergalactic magnetic
fields are assumed to be responsible for these energy-dependent
deviations from the direction to M87.

This scenario predicts  a strong asymmetry between the
north and south galactic hemispheres.
Consider for instance that all EHECRs enter 
the galactic halo less than $15^\circ$ away from  the
direction to the north galactic pole. The lines in 
Figure~3 display the southernmost  possible arrival
directions on Earth, for several values of $E/Z$. 

\FIGURE{\epsfig{file=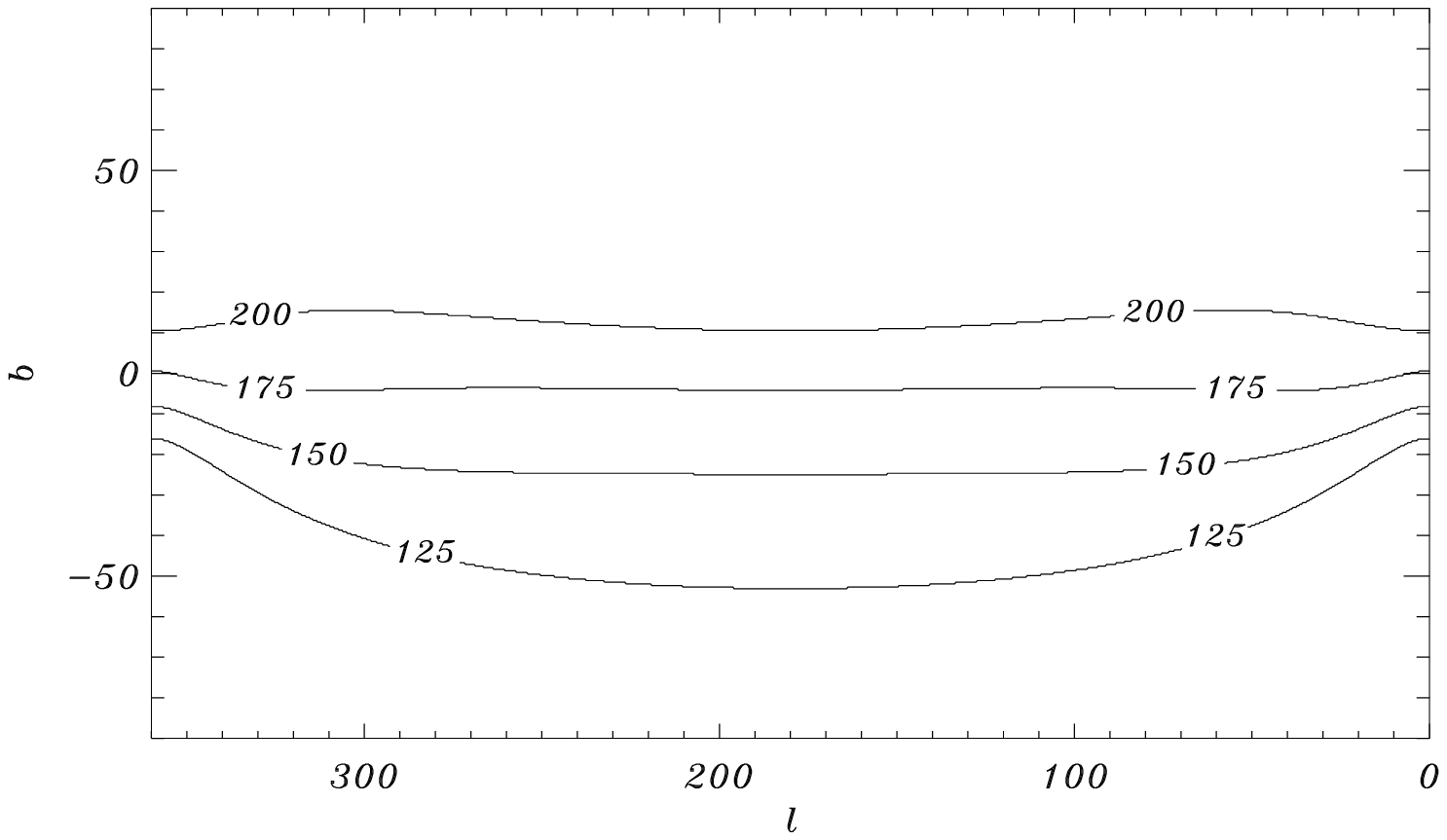,width=16cm}
\caption{Cosmic rays that enter the galactic wind less than
$15^\circ$ away from the direction to the north galactic pole
reach Earth from directions above the lines drawn,
corresponding to different values of $E/Z$.}\label{cutoff}}

This scenario thus implies that no charged cosmic rays
should arrive on Earth from directions below the lines drawn 
in Fig.~3 for values of $E/Z$ larger than that indicated along 
each line. Certainly CRs may arrive below these lines if they 
enter the galactic wind with an inclination larger than $15^\circ$ 
from its symmetry axis, but then they do so with smaller 
magnifications (or rather with large demagnifications) as 
larger is the inclination. Thus, even if there were other 
EHECR sources as powerful as M87 in our local (less than 100~Mpc)
neighborhood, their flux will not be significantly magnified 
(or rather they will be significantly demagnified) if the CRs enter
the galactic wind far away from the direction to the north 
galactic pole.  Only if $E/Z$ is above a few times $10^{21}$~eV
does the observed flux approach its unlensed value, and the CRs 
arrival directions point to their true source location. 

Even if there is a unique source, 
CRs at different energies should enter the galactic wind 
from different directions if they have suffered magnetic deflections
in their way. It is nevertheless instructive to
consider fixed incoming
directions, to further illustrate some of the generic features of  
lensing by the galactic wind already discussed in the 
previous sections. 
Figure~4 displays the energy-dependent magnification of the
flux of CRs that enter the galactic magnetic wind from directions
($\ell, b)=(270^\circ,88^\circ)$ and $(270^\circ, 85^\circ)$, 
for the principal (P) as
well as for the secondary images (A, B). 
Figure~5 displays the change in the observed arrival directions of
CRs with these entrance directions as the energy is lowered  
down to $E/Z=75$~EeV.

\FIGURE{\epsfig{file=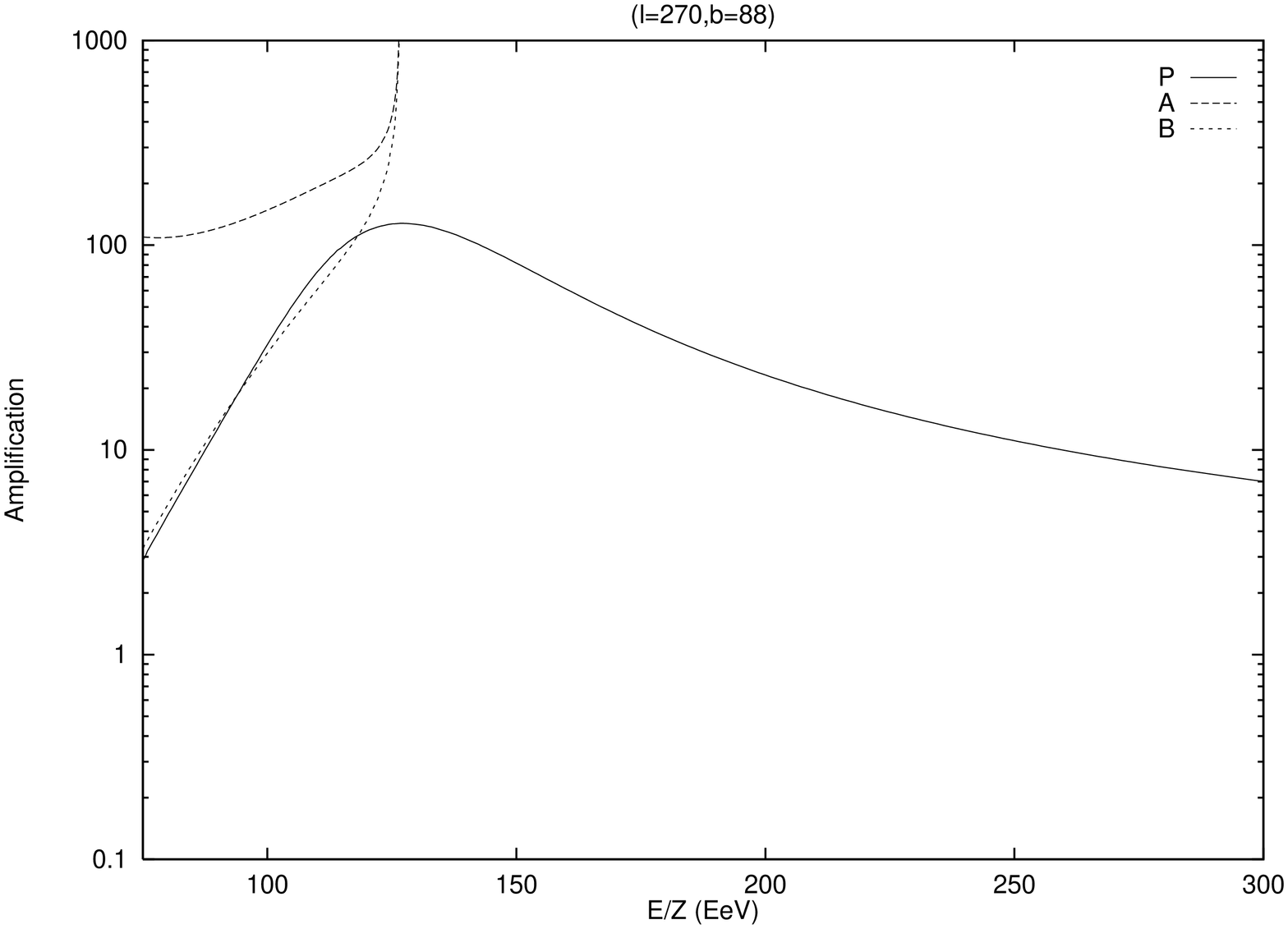,width=5cm,angle=-90}
\epsfig{file=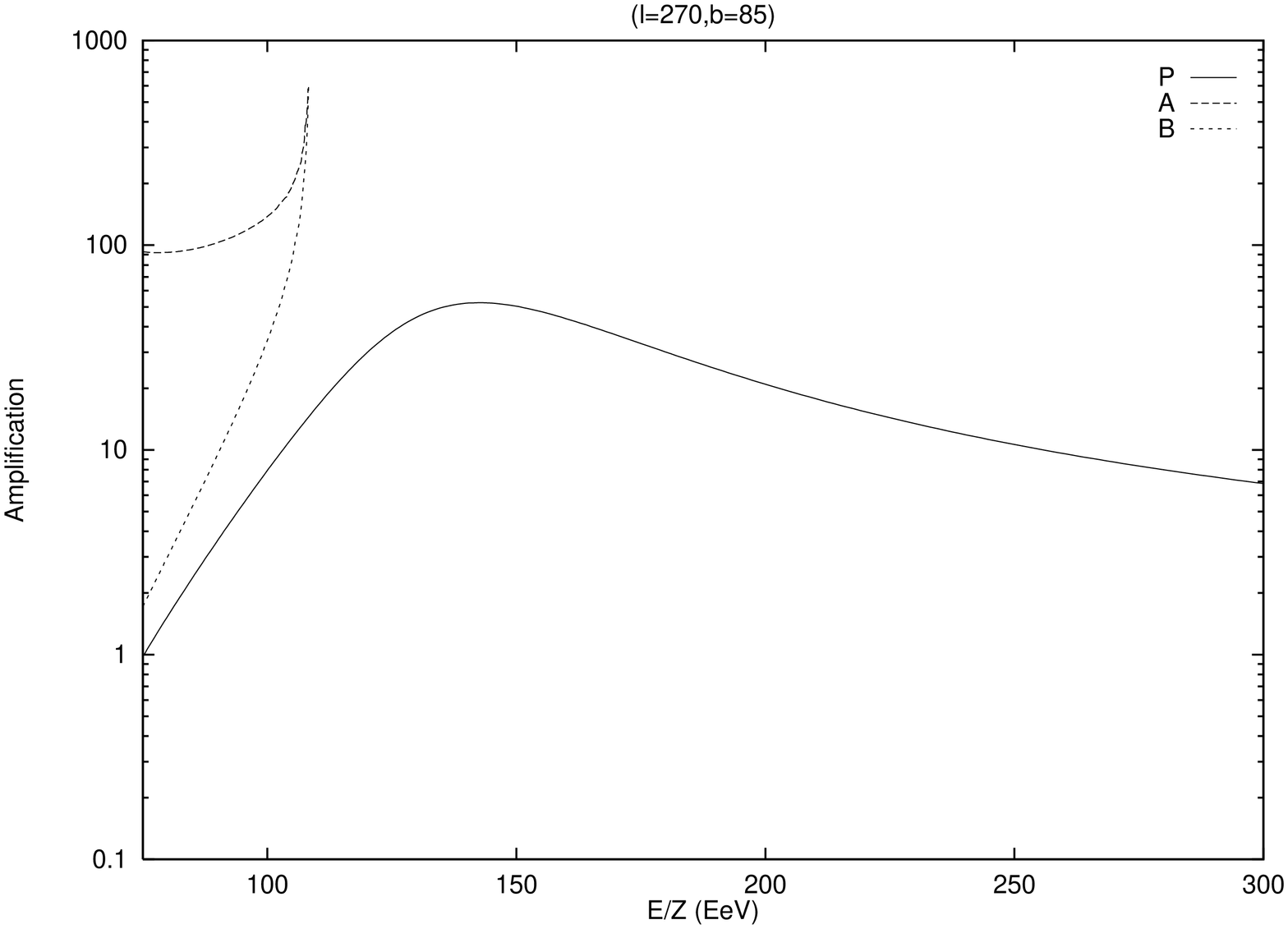,width=5cm,angle=-90}
\caption{Energy-dependent amplification of the CR flux 
that enters the galactic wind from $(\ell,b)=(270^\circ,88^\circ)$
(left) and $(\ell,b)=(270^\circ,85^\circ)$
(right).}}\label{spectra}

\FIGURE{\epsfig{file=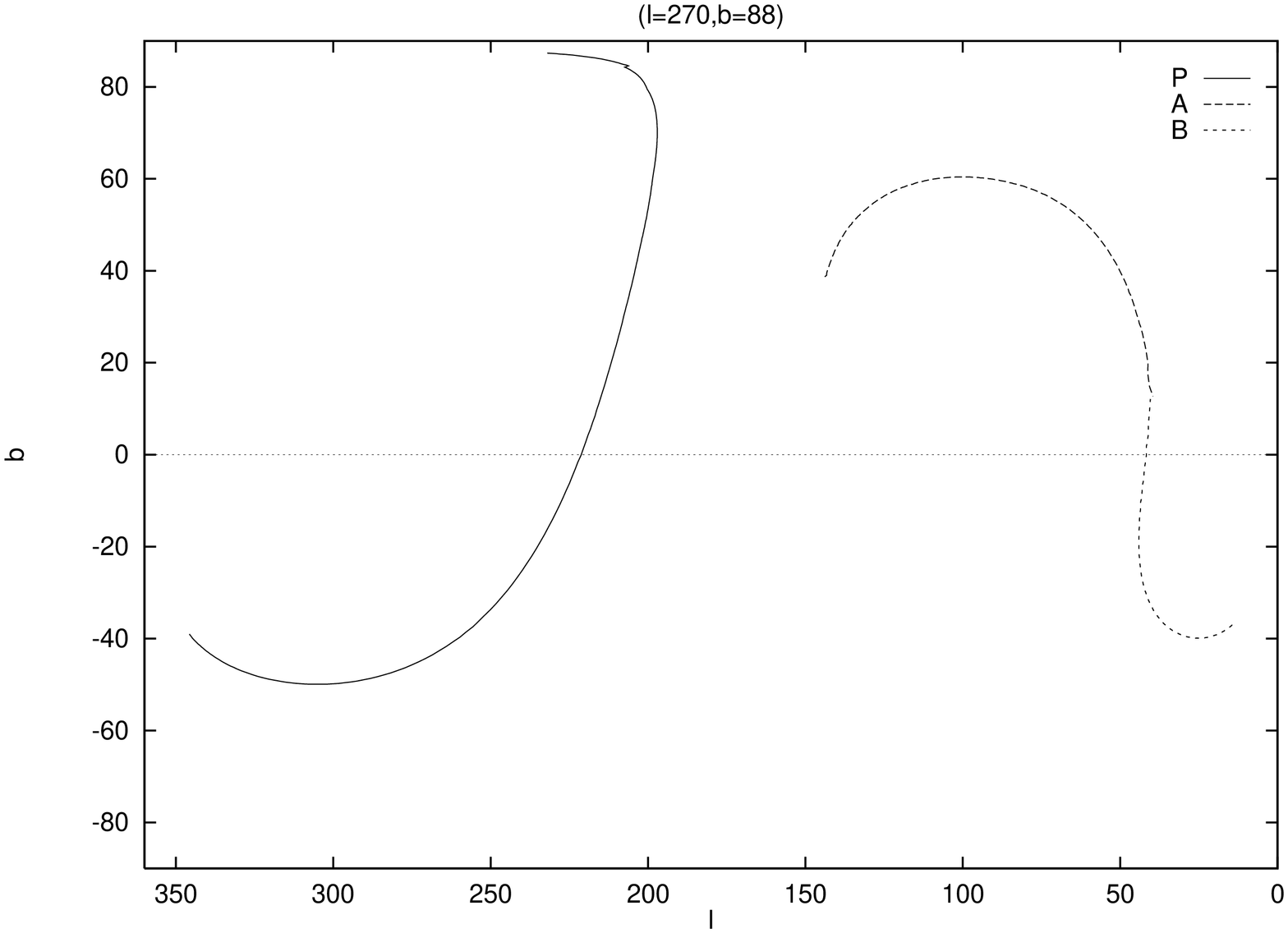,width=5cm,angle=-90}
\epsfig{file=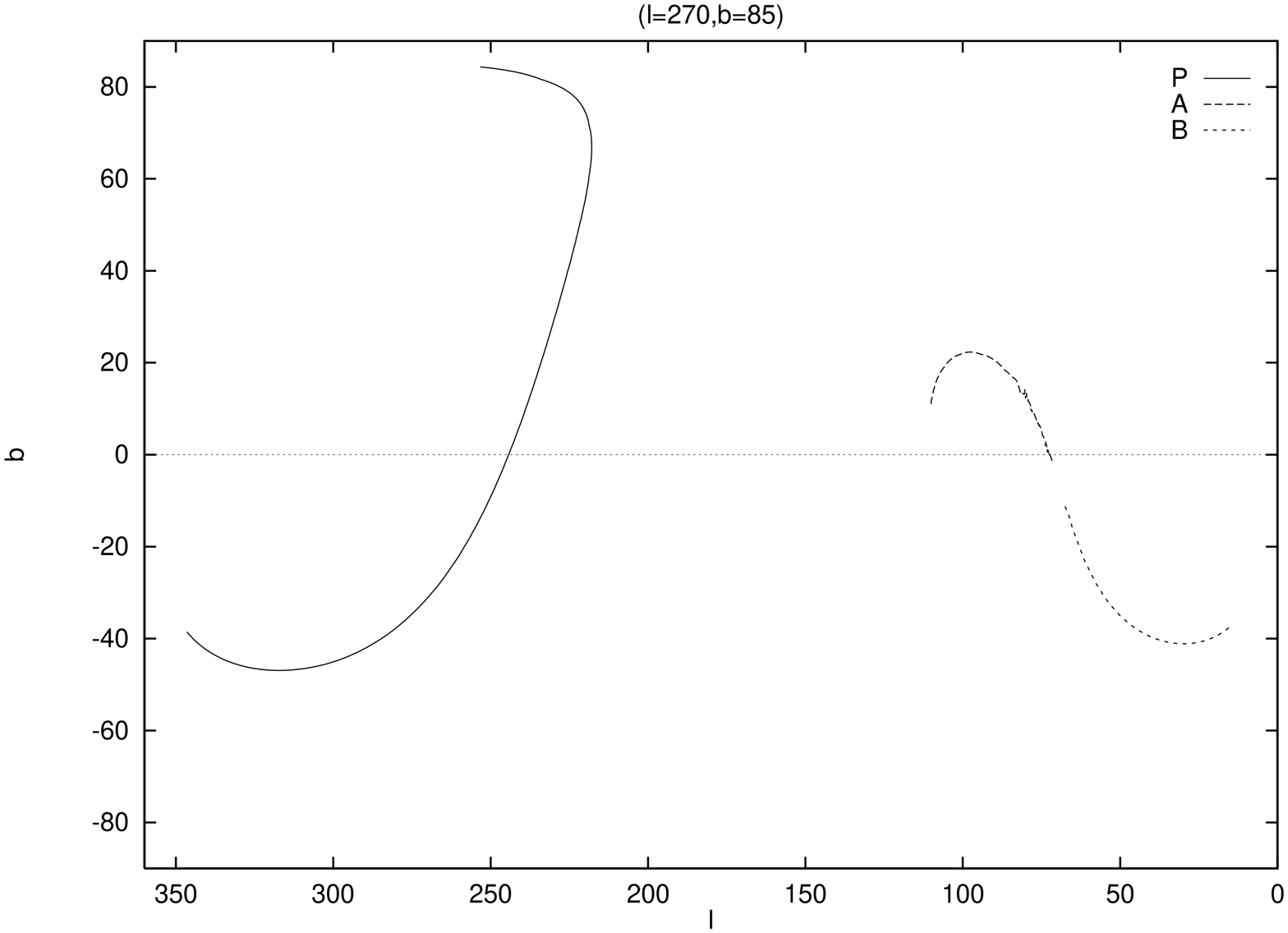,width=5cm,angle=-90}
\caption{Arrival directions of CRs that enter the galactic 
 wind from $(\ell,b)=(270^\circ,88^\circ)$
(left) and $(\ell,b)=(270^\circ,85^\circ)$
(right), as their energy falls down to $E/Z=75$~EeV}}
\label{displacements}

The principal image is magnified by a factor of order 10 at 
$E/Z\approx 200$~EeV, is further amplified at intermediate
energies, and then its magnification starts to rapidly decrease
while $E/Z$ is still above 100~EeV. Its apparent position moves
south as the energy decreases. 
Secondary images appear at the energy at which the caustic
crosses the source position, with formally divergent magnifications. 
One of the secondary images moves north and remains highly magnified, 
with an amplification factor above 100, while the other moves south
and is quickly demagnified. The secondary images appear and remain 
in the  opposite east-west hemisphere than that in which the principal 
image is seen.

The general features displayed in these examples are generic
to different entrance directions. However, the precise 
energy and location at which secondary images form, the energy 
at which the principal image is most magnified, and the exact 
amount of magnification attained depend sensibly on the entrance
direction. For instance, if the entrance direction is more than 
about $8^\circ$ away from the north galactic pole, 
then  secondary images appear only below 100~EeV, 
and the principal image acquires smaller
and smaller maximum magnification as the entrance direction is 
farther away from the vertical. 

The expected energy and angular distribution of observed EHECRs 
can be exemplified considering again some fixed 
entrance directions and assuming that  the differential
flux injected by the source scales for instance as $E^{-2.7}$. 
We take the detecting system to have the same efficiency at all energies 
within the range considered, which we divide in 50 bins of 
equal detection probability. We consider values of $E/Z$ larger
than 100~EeV only. Figure~5 displays
the arrival directions of the events that would be 
detected in each case.

\FIGURE{\epsfig{file=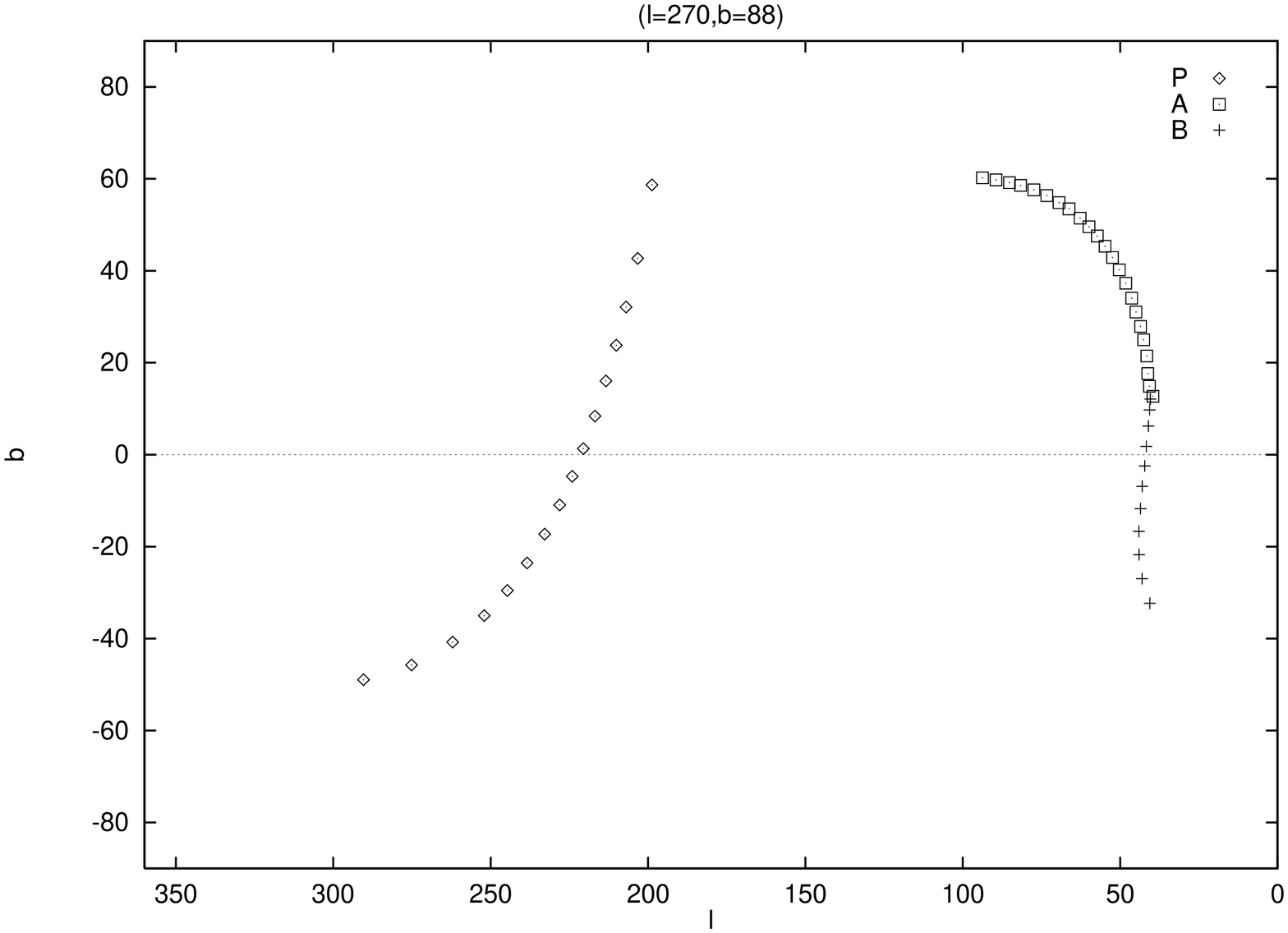,width=5cm,angle=-90}
\epsfig{file=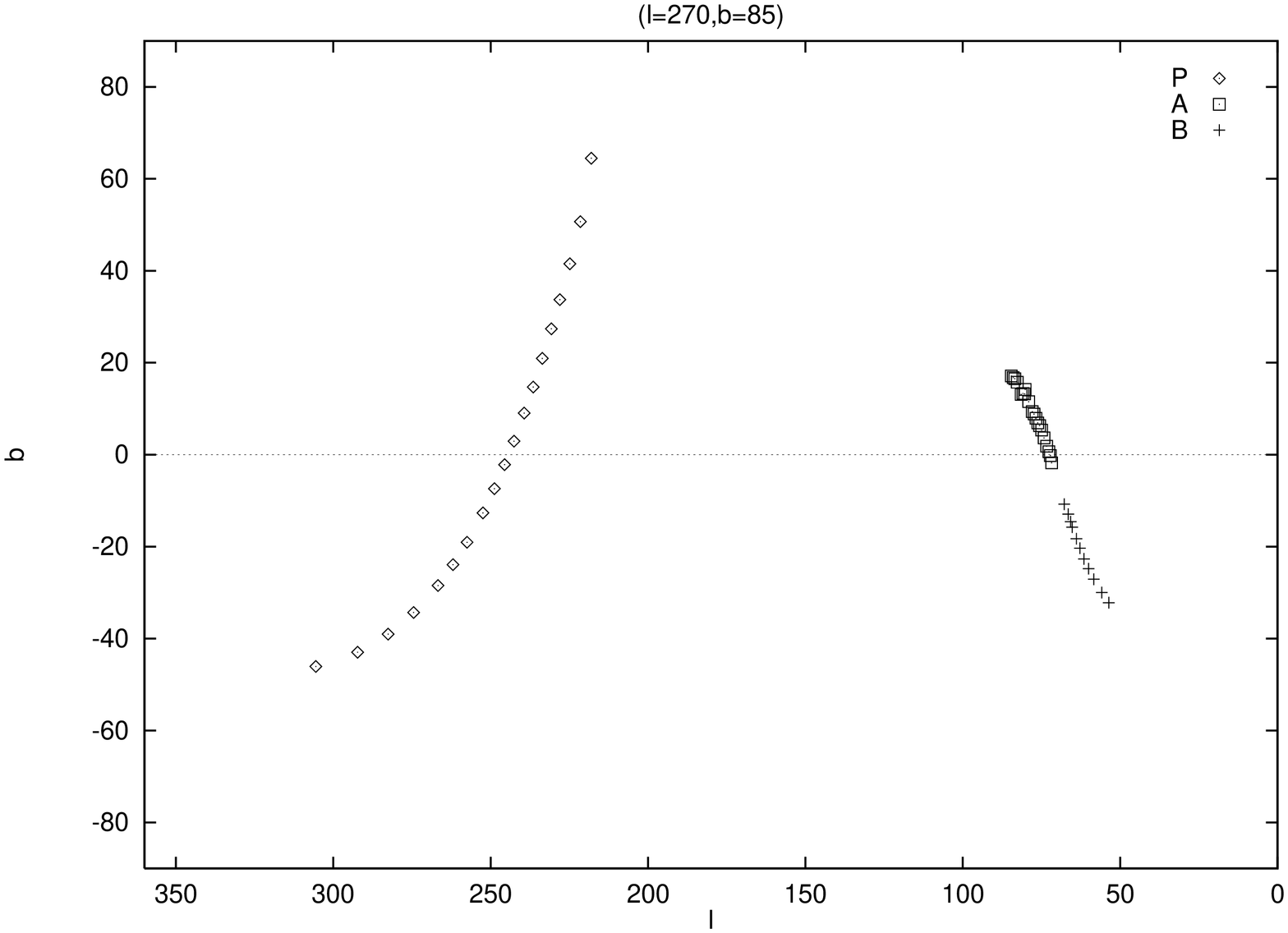,width=5cm,angle=-90}
\caption{Predicted arrival directions of 50 EHECR events with
$E/Z$ above 100~EeV, if they enter the galactic wind from 
$(\ell,b)=(270^\circ,88^\circ)$ (left) and  $(\ell,b)=
(270^\circ,85^\circ)$ (right), assuming an injection flux proportional
to $E^{-2.7}.$}\label{ev}}

As already discussed, we see that
 there are no events at southern galactic
latitudes below a certain energy threshold. 
Indeed, the apparent position
of the principal image crosses the galactic equator at
$E/Z\approx 133$~EeV in the left panel, 
$E/Z\approx 144$~EeV in the right, and 
the events in the secondary images are all below the energy 
of the caustic ($E/Z\approx 127$~EeV in the left panel, 
$E/Z\approx 108$~EeV in the right).  Notice also that the divergence
 in the magnification at the caustic gives anyhow a finite number of
 events once it is convoluted with the differential spectrum
of the incident CRs \cite{II}.

\section{Conclusions}\label{conclusions}

We have analysed flux magnification and multiple image
formation  in the galactic magnetic wind scenario put forward 
in \cite{ahn99}. In this scenario all
observed arrival directions of EHECRs are
compatible with a common origin in M87 if 
intergalactic magnetic fields provide the extra deflection 
(of order $20^\circ$) needed to fine-tune the incoming particles 
in the appropriate direction as they enter the wind.
We find that magnification factors well above 100
are attained in a significant energy range, with $E/Z$ below
 150~EeV. This reduces the
energy requirements upon the source, that would need to be a factor of
more than
100 less powerful than if unlensed to provide the same 
observed flux in this energy range.

One of the definite predictions of this model is the strong asymmetry
expected between events arriving from the northern and southern
galactic hemispheres. Although with the present EHE data, which
involves only the northern terrestrial hemisphere and hence mainly the
northern galactic one, it is not yet possible to test this
asymmetry,
the future operation of the Auger observatory, that will provide good
coverage of the southern skies, will allow to check the viability of
this model. In particular, a very strong suppression of events above
$E\simeq 150$~EeV should be present at latitudes below $b\simeq -30^\circ$
for the scenario to survive.
Another general feature is that an abrupt kink in the
overall spectrum should appear when the secondary images disappear,
i.e. when the energy increases beyond the
energy of the caustic crossing. Although no particular feature of this
kind is apparent in the present data, 
an increased statistics in the northern
skies would be desireable 
to definitely confront this  prediction with observations.

As a final remark, we would like to point out that if a galactic wind
is indeed present in the Milky Way, but with a smaller overall
strength (so that for instance locally it is below the 2--3 $\mu$G
amplitude of the spiral field which is inferred from observations,
rather than the 7 $\mu$G adopted here), it could in any case have
interesting observational consequences, especially if EHECRs have a
component which is not light. In particular, one can think of a
scenario in which an extragalactic source near the north pole, such as
M87, produces heavy nuclei with $E>10^{20}$~eV (and not necessarily
protons at these energies), and their flux is strongly amplified by the
galactic wind field. In this case a transition to a heavy composition 
could result at extremely high energies \footnote{One would also have 
to take
into account here that for propagation distances beyond $~\sim 10$~Mpc the
photodisintegration of nuclei out of the CMB photons 
starts to be important 
for $E>Z\times 10^{19}$~eV \cite{pu76}.}.

\acknowledgments

Work partially supported by ANPCyT, CONICET and Fundaci\'on Antorchas, 
Argentina.

\end{document}